# MatChat: A Large Language Model and Application Service Platform for Materials Science

Ziyi Chen[1,2,†]，Fankai Xie[3,4,†]，Meng Wan[1,†],Yang Yuan[1,2], Miao Liu[3,5,6,*], Zongguo Wang[1,2,*], Sheng Meng[3,5], Yangang Wang[1,2]

[1]Computer Network Information Center, Chinese Academy of Sciences, Beijing, 100083, China

[2]University of Chinese Academy of Sciences, Beijing, 100049, China

[3]Institute of Physics, Chinese Academy of Sciences, Beijing 100190, China

[4]School of Physical Sciences, University of Chinese Academy of Sciences, Beijing, 100190, China

[5]Songshan Lake Materials Laboratory, Dongguan, Guangdong, 523808, China

[6]Center of Materials Science and Optoelectronics Engineering, University of Chinese Academy of Sciences, Beijing, 100049, China

*Corresponding authors: Miao Liu(mliu@iphy.ac.cn),

Zongguo Wang(wangzg@cnic.cn)

[†]These authors contributed equally to this work

## Abstract

The prediction of chemical synthesis pathways plays a pivotal role in materials science research. Challenges, such as the complexity of synthesis pathways and the lack of comprehensive datasets, currently hinder our ability to predict these chemical processes accurately. However, recent advancements in generative artificial intelligence (GAI), including automated text generation and question-answering systems, coupled with fine-tuning techniques, have facilitated the deployment of large-scale AI models tailored to specific domains. In this study, we harness the power of the LLaMA2-7B model and enhance it through a learning process that incorporates 13,878 pieces of structured material knowledge data. This specialized AI model, named MatChat, focuses on predicting inorganic material synthesis pathways. MatChat exhibits remarkable proficiency in generating and reasoning with knowledge in materials science. Although MatChat requires further refinement to meet the diverse material design needs, this research undeniably highlights its impressive reasoning capabilities and innovative potential in the field of materials science. MatChat is now accessible online and open for use, with both the model and its application framework available as open source. This study establishes a robust foundation for collaborative innovation in the integration of generative AI in materials science.

## 1. Introduction

At present, Large Language Models (LLMs) have established a robust foundation for various applications. OpenAI's ChatGPT and GPT-4.0[1], with 175 billion and 18 trillion parameters respectively, clearly represent a new era in the development of artificial intelligence (AI). However, OpenAI has not disclosed the specific details of the training methods and parameters of the model. Tsinghua's GLM base model [2-3] provides a compelling option for natural language processing. It supports both English and Chinese, offering high accuracy, cross-platform compatibility, reproducibility, and fast inference. Baidu's Ernie 3.0 Titan, an evolution of the Ernie series models [4-6] with an impressive 260 billion parameters, stands as the largest Chinese dense pre-training model to date, with great potential for deep language understanding and applications. The LLaMA and LLaMA2 models [7-8], ranging from 7 billion to 70 billion parameters, contribute to the diversity of open-source large language models, catering to various applications. The Ziya-LLaMA-13B pre-training model [9] exhibits robust capabilities across domains such as translation, programming, text classification, information extraction, summarization, copywriting, common sense question answering, and mathematical computation. The outstanding performance of these models offers strong support for various tasks and holds the promise of unlocking potential in other domains.

Fine-tuning open-source large models has emerged as an effective method for tailoring AI capabilities to meet the specific demands of various domains. Currently, fine-tuning techniques have demonstrated considerable success in vertical fields, including healthcare, education, and finance. In the field of healthcare, models like HuatuoGPT[10] and DoctorGLM[11] have been developed to address medical challenges, these models exhibit a high degree of professionalism and offer invaluable insights within the healthcare domain. In the finance sector, notable strides have been made with the XuanYuan[12] model, its application has brought substantial benefits and advancements to financial operations. Similarly, in the education sector, the EduChat[13] model has demonstrated its worth by delivering valuable capabilities tailored to educational contexts. Additionally, the Fengshenbang[14] large model system, a product of the Cognitive Computing and Natural Language Research Center at IDEA Institute, has gained widespread recognition. The Fengshenbang large model system is a Chinese language-centric ecosystem that includes pre-training of large models and fine-tuned applications tailored for specific tasks, benchmarks, and datasets. Its overarching objective is to create a comprehensive, standardized, and user-centric ecosystem.

In recent years, researchers have achieved significant and innovation results in the discovery of new materials [15] and their theoretical interpretation [16] by leveraging the existing database such as Atomly[17-18], OQMD[19], MaterialsProject[20], and others. They have successfully explored the intricate relationships between material structure and properties [21], addressing the challenges posed by the scarcity of materials data through the development of more accuracy artificial intelligence optimization [22] and training methods [23]. With the application of large models, researchers in materials science have explored the use of these models to tackle challenges such as the intricate nature of chemical reactions and structures. One notable example is the MatSciBERT[24] model which is derived from BERT[25]. MatSciBERT exhibits the capability to automatically extract information from literature, conduct data mining, and construct knowledge graphs, thereby ushering in new possibilities for the application of language models in the field of materials science. To the best of our knowledge, there has been no reported utilization of large language models in material science until now.

To advance the innovative application of large language models in the field of materials science, this study employs a carefully constructed knowledge question-answering database to explore their potential in materials science. We propose a viable solution for predicting inorganic material chemical synthesis pathways and provide a preliminary demonstration of the feasibility of this approach. To optimize the performance of the large model in answering questions related to material synthesis knowledge, our research adopts the LLaMA2-7B model as a pre-training model. This approach involves a combination of supervised fine-tuning and reinforcement learning, incorporating valuable human feedback to enhance model optimization. The dataset selected for this purpose comprises 35,675 solution-based synthesis processes [26] extracted from scientific papers. Following thorough processing, we obtained a dataset consisting of 13,878 high-confidence synthesis pathway descriptions.

Although the relatively modest model parameters used in this study, result in cost-effective training, the model has showcased impressive comprehensive reasoning abilities.

The highlights of this study include two primary aspects. (1). Fine-tuning the LLaMA2-7B pre-training model using the preprocessed dataset of inorganic material synthesis program instruction. (2). Development of a question-answering platform for the materials synthesis large language model, aimed at facilitating work within the field of materials science and providing an accessible and user-friendly dialogue platform. This paper's basic structure comprises the following sections. Section 2 focuses on the details of the model fine-tuning process. In section 3, we explore the construction of the question-answering platform, covering aspects such as architecture design, parallel processing, resource management, and other technologies. Section 4 presents the experimental findings, and section 5 serves as the conclusion of this study.

## 2. Fine-tune MatChat Model Methods

### 2.1 Base Model

LLaMA2, an updated iteration of LLaMA1, has been trained by Hugo's team [8] on a revised combination of publicly available datasets. The pretraining corpus size has been increased by 40\%, the model's context length has been doubled, and a grouped-query attention mechanism has been adopted. Variants of LLaMA2 with 7B, 13B, and 70B parameters are being released to the public. Based on the results of the paper, both LLaMA2 7B and 30B models outperform MPT models of equivalent sizes in all categories [8].

The model in our work is fine-tuned based on the open-source large language model, LLaMA2-7B, which has 7 billion parameters, a content length of 4k, and supports up to 2.0 trillion tokens.

## 2.2 Materials Knowledge Data

The dataset used for fine-tuning the model in this paper is derived from 35,675 solid-phase synthesis processes of inorganic materials extracted from over four million papers. After rigorous screening, deduplication, and cleaning, we obtained a training set consisting of 13,878 highly reliable synthesis pathway descriptions. This dataset was further preprocessed and integrated into an instruction question-answering format, as shown in Figure 1. The prompts involve specific material synthesis method inquiries, and the responses provide the corresponding chemical reactions and synthesis conditions.

```
{
  "instruction": "How to synthesize Li4Ti5O12",
  "input": "How to synthesize Li4Ti5O12",
  "output": "Li4Ti5O12 can be made from TiO2, and Li2CO3 via solid state
            reaction. The chemical reaction equation is (2 Li2CO3 + 5 TiO2
            == 1 Li4Ti5O12 + 2 CO2). The detailed recipe can be found in
            literature [10.1149/1.1383553]."
}
```

Figure.1 The instruction format for the question-answering scenario

## 2.3 Training Process

The model fine-tuning process utilized the following parameters, a learning rate of $10^{-4}$, a batch size of 8, and one epoch for fine-tuning. All fine-tuning operations were executed on NVIDIA A100 GPUs.

While fine-tuning the LLaMA2 model, we used two methods and respective resource management strategies. Firstly, the "Parameter Efficient Model Fine-Tuning" approach aimed to make fine-tuning economically feasible on a single consumer-grade GPU. This method involved freezing the entire model and adding small learnable parameters or layers, training only a fraction of the model's parameters. Methods like LORA, LLaMA Adapter, and Prefix-tuning were employed, addressing cost, deployment, and avoiding catastrophic forgetting.

Alternatively, the "Full/Partial Parameter Fine-Tuning" method offered flexibility. We could freeze most pre-trained model layers and fine-tune only the task-specific head, add extra fully connected layers, or fine-tune all layers. For larger models, multiple GPUs might be required, especially when a single GPU couldn't accommodate the model.

To tackle multi-GPU training challenges, we used the "Fully Sharded Data Parallel" (FSDP) technique as noted on the GitHub Repository [27]. FSDP shards data, model parameters, gradients, and optimizer states across GPUs, saving memory and enabling larger models on the same number of GPUs.

## 3. MatChat Platform

To support researchers in obtaining fast and accurate model inference results, we have developed a set of web-based dialogue service interfaces based on LLaMA2. This section focuses on explaining how to construct these service interfaces, including the associated technical details and implementation methods.

### 3.1 Architecture and Method Design

In the development of the MatChat platform, we employed PyTorch as the core computing framework to handle tasks such as loading, running, and reasoning with large models. For the web service interface, we chose Python Flask to manage both HTTP and WebSocket requests, facilitating seamless integration with PyTorch. SocketIO was implemented for efficient, event-based two-way communication. When users request model reasoning, SocketIO delivers the model's output in real-time, eliminating traditional polling delays. Flask is responsible for handling user HTTP requests, parsing input parameters, and scheduling background model runs.

To ensure rapid user authentication and system stability, we implemented lightweight data storage in Redis for token verification and resource isolation during concurrent usage. Redis, as an in-memory data structure storage, offers fast read and write capabilities, making it suitable for high-concurrency scenarios. Furthermore, Redis-based token verification enhances system security. When a user submits a request, the system queries Redis to validate tokens, thereby enhancing security against potential malicious activity.

### 3.2 Concurrency Processing and Resource Management Technologies

In scenarios with concurrent access from multiple users, efficient resource management becomes crucial. To address resource contention, we implemented a waiting queue based on condition variables. This design offers several advantages as follows:

(1) Automatic entry into waiting state: In situations where resources are occupied, new requests seamlessly transition into the waiting state.

(2) Sequential awakening of queued requests: Upon resource release, requests within the waiting queue are sequentially awakened, allowing them to acquire the resources.

(3) Thread locks for exclusive access: Thread locks guarantee exclusive resource access for a single request at any given time, mitigating potential data competition issues.

This mechanism ensures the system's functions to consistently provide services to each user, even in a high-concurrency environment, maintaining stability throughout.

### 3.3 Deployment and optimization of LLaMA2 model

As a deep learning model, the deployment of the LLaMA2 model presents a myriad of challenges, including high computing resource requirements, a complex model structure, a substantial number of parameters, and extensive demands on memory and processing power. To meet the need for real-time user responses, the model must exhibit swift inference capabilities.

We devised a mode employing half-precision floating-point numbers (float16) for loading the model. This approach significantly reduced both memory usage and computation time. Additionally, we leveraged PyTorch's compile function to further optimize the model's runtime efficiency. Furthermore, we implemented a streaming output feature for the model, allowing users to observe results in real time during the model's execution, thereby enhancing the user experience.

Considering the intricacy and computational demands of the LLaMA2 model, we introduced a resource scheduling mechanism to ensure seamless responses for concurrent users. When a user requests model resources, the system assesses resource availability by competing for locks. If GPU resources are occupied and in the inference state, the user's request is placed in a waiting queue, persisting until the resources become available. Through this mechanism, the system guarantees that only one request accesses the model at any given time, mitigating potential resource contention issues. Conversely, when a user obtains the lock resource and initiates inference, the streaming output doesn't wait for the entire sequence to complete. Instead, it continues to generate and dispatch results in real time.

## 4. Experiment

### 4.1 Baseline

In the experimental stage, given the lack of large models specifically tailored for inorganic material synthesis knowledge question-answering, we opted for the widely-used general large models-ChatGPT, Ernie Bot, Spark Desk, ChatGLM-for a comparative experiment on the performance of inorganic materials synthesis question-answering. Details can be found in Table 1 for information. Among them, the information on the Spark Desk model is not disclosed.

Table.1 Model information for experimental comparison]

| Model | Parameters | Base Model |
|-------|-----------|-----------|
| ChatGPT | 175B | gpt-3.5-turbo |
| Ernie Bot | 260B | Ernie 3.0 Titan |
| Spark Desk | - | - |
| ChatGLM | 6B | GLM-130B |

4.2 Metrics

When evaluating natural language processing models, a comprehensive assessment often involves a combination of BLEU and ROUGE metrics. BLEU primarily measures the accuracy and exact matching of translation, with an emphasis on Precision, while ROUGE evaluates information completeness and coverage in summaries, emphasizing Recall.

However, when dealing with extensive language models in the domain of inorganic material synthesis question answering, our primary focus is on observing the safety, accuracy, and usability of the generated answers—a metric we refer to as SAU. Safety involves ensuring that the resulting material synthesis process does not pose potential dangers or can highlight possible hazards. Accuracy requires that the generated answers be as precise as possible, factoring in crucial elements such as required raw materials, time, and temperature. Usability emphasizes that the model's answers should be highly specific, encompassing operational details like the synthesis environment and utilized devices.

Furthermore, we demonstrated the generative inference capability of our model, showcasing its proficiency in providing synthesis processes for structures not present in the training set.

4.3 Results

We conducted a comprehensive performance evaluation, comparing our model against several baseline models. The questions posed were primarily skewed towards probing the knowledge of inorganic material synthesis processes, with an emphasis on extracting synthesis methods mentioned in the literature [28-29]. To enhance clarity, we extracted key synthetic raw material contents from model answers, as illustrated in Table 2. Identical questions were input to both our model and the baseline models during the experiment, and the corresponding output responses were observed. To maintain brevity, we condensed lengthy model answers by extracting essential synthetic raw material details. Furthermore, our input questions were rooted in relevant synthesis literature, utilizing chemical formulas such as $K_2CaMoB_2P_2O_{13}$[28] and $Nd_2Zn_{60}B_{40}Si_{19}O_{161}$[29].

We first delve into the analysis of the answer regarding $K_2CaMoB_2P_2O_{13}$. In terms of safety, all models perform similarly. Concerning accuracy, both ChatGPT and Spark Desk provide answers, but the raw materials mentioned in their responses are found to be incorrect based on relevant literature. Ernie Bot and ChatGLM models fail to furnish answers. Notably, our MatChat model not only provides an answer but also presents synthetic raw materials that closely aligned with those detailed in the literature. Moreover, our model outshines others in terms of usability by offering the most informative responses.

Then, turning our attention to the answers concerning $Nd_2Zn_{60}B_{40}Si_{19}O_{161}$, the models demonstrate comparable performance in terms of safety. However, in terms of accuracy, ChatGPT and ChatGLM models provide vague raw material information for various elements, lacking practical guidance. The Spark Desk model offers guidance in the form of oxides for each element, but the literature indicates that the source of the B element is H3BO3. Ernie Bot fails to provide a relevant answer. On the other hand, our MatChat model delivers raw material information closest to the literature, showcasing the highest guiding value.

In summary, MatChat proves to be highly valuable in predicting material synthesis processes, particularly for its accuracy and usability.

Table.2 Experiments of model comparison

| | How to make $K_2CaMoB_2P_2O_{13}$? | How to make $Nd_2Zn_{60}B_{40}Si_{19}O_{161}$? |
|---|---|---|
| ChatGPT | Materials Needed:<br>1. Potassium carbonate (K2CO3)<br>2. Calcium oxide (CaO)<br>3.Ammonium molybdate (($NH_4)_6Mo_7O_{24}\cdot4H_2O$)<br>4. Boric acid ($H_3BO_3$)<br>5. Phosphoric acid ($H_3PO_4$)<br>6. Deionized water<br>7. Glassware (flasks, beakers, etc.)<br>8. Heating equipment (hot plates, ovens, etc.)<br>9. Laboratory equipment (stirrers, pipettes, etc.)<br>10. Protective gear (lab coat, gloves, safety goggles, etc.) | Gather the Starting Materials:<br>You'll need the following starting materials, which are the elements and compounds that make up the compound you want to synthesize:<br>1. Neodymium (Nd) compounds<br>2. Zinc (Zn) compounds<br>3. Boron (B) compounds<br>4. Silicon (Si) compounds<br>5. Oxygen (O) sources (e.g., oxygen gas, oxides) |
| Ernie Bot | As an artificial intelligence language model, I cannot provide information about the production of chemical substances. | $Nd_2Zn_{60}B_{40}Si_{19}O_{161}$ is a complex oxide ceramic material with a formula derived from the formula of the lanthanide series. It is a composite oxide ceramic material that has not been widely commercialized and is difficult to produce. The production process is complex and requires high technology. In this case, the best advice is to seek help from professionals. |
| Spark Desk | $K_2CaMoB_2P_2O_{13}$ is a chemical compound with the formula $K_2CaMoB_2P_2O_{13}$. It is a type of potassium phosphate, commonly known as plaster of Paris. Making | To make $Nd_2Zn_{60}B_{40}Si_{19}O_{161}$, one would need to start with the raw materials: $Nd_2O_3$, ZnO, $B_2O_3$, $SiO_2$, and $H_2O$. The specific ratios of these components would need to be |
| | plaster of Paris involves a process of mixing water and calcium hydroxide ($Ca(OH)_2$) until it forms a thick paste. | carefully controlled during the synthesis process to achieve the desired composition. |
| ChatGLM | I'm sorry, but I cannot provide a specific method for creating the chemical compound $K_2CaMoB_2P_2O_{13}$, as it is not a commonly known or widely used compound. | $Nd_2Zn_{60}B_{40}Si_{19}O_{161}$ is a chemical formula for a specific alloy, which is a combination of neodymium (Nd), zinc (Zn), boron (B), silicon (Si), and oxygen (O). To make this alloy, you would typically start with the appropriate elements in their elemental forms and then mix them in the correct stoichiometric ratio. |
| MatChat | $K_2CaMoB_2P_2O_{13}$ can be made from $K_2CO3$, $(NH_4)_2HPO_4$, $CaCO_3$, $H_3BO_3$, and $MoO_3$ via solid state reaction. The chemical reaction equation is ($2(NH_4)_2HPO_4$ + $1CaCO_3$ + $1H_3BO_3$ + $1K_2CO_3$ + $0.5MoO_3$ == $1K_2CaMoB_2P_2O_{13}$ + $2CO_2$ + $3H_2O$ + $4NH_3$). The precursors are ground in agate, heated under 600.0°C for 24.0h, cooled. The detailed recipe can be found in literature [10.1016/s0167-2738(03)00122-0]. | $Nd_2Zn_{60}B_{40}Si_{19}O_{161}$ can be made from $SiO_2$, $Nd_2O_3$, ZnO, and $H_3BO_3$ via solid state reaction. The chemical reaction equation is ($40H_3BO_3$ + $19SiO_2$ + $1Nd_2O_3$ + $60ZnO$ == $1Nd_2Zn_{60}B_{40}Si_{19}O_{161}$ + $60H_2O$). The precursors are mixed in agate, heated under 1000.0°C in air for 24.0h, cooled. The detailed recipe can be found in literature [10.1016/j.jssc.2005.06.002]. |
| Litera ture | The studied glasses have the composition expressed by the formula $xMoO_3 \cdot (1-x)$ $[48P_2O_5 \cdot 45CaO \cdot 2B_2O_3 \cdot 5K_2O]$ with x = 0, 1, 3, 5 and 7 mol%. They were prepared using the conventional melt quenching method. Appropriate quantities of reagent grade $MoO_3$, $NH_4 \cdot H_2PO_4$, $CaCO_3$, $H_3BO_3$ and $K_2CO_3$ were mixed in an agate mortar. The batches were melted in air, in sintered corundum crucibles, in an electric furnace at 1100 °C for 25 min. The melts were quickly cooled at room temperature | Zinc-borosilicate glasses in the chemical composition of 60ZnO–20B_2O_3–19SiO_2–1Nd_2O_3(mol%) have been prepared by employing a conventional quenching method from the spectral pure grade raw chemicals such as ZnO, $SiO_2$, $H_3BO_3$ and $Nd_2O_3$. The batches of chemical mix weighing about 50 g were melted in a platinum crucible at 1300 °C for about 3 h in computer-controlled electrical furnace. The melts were poured onto a smooth surfaced stainless steel plate and pressed with another similar plate in order to obtain a few circular glass |

| | by pouring and stamping between two copper plates previously cooled with liquid nitrogen. The glass samples were ground to powder in a Retsch Planetary ball mills, type PM 100. The average size of the obtained grains was about 30 μm [28]. | discs of 2–3 cm in diameter with a thickness of 0.3 cm each. These samples were annealed at 550°C for 1 h and cooled down slowly to the room temperature to remove internal stresses present in the glass samples [29] |
|---|---|---|

Furthermore, we showcase the dual capabilities of our model, encompassing both generative and inferential aspects. Our training set comprises a total of 13,878 diverse chemical formula synthesis data. When we query the model using chemical formulas present in the dataset, the output exhibits a degree of inconsistency with the training set data, highlighting the model's generalization capabilities. Moreover, when posing questions with chemical formulas absent from the dataset, the output format and content align in structure with the dataset, offering valuable insights for the synthesis process.

## 5. Conclusion

Based on the LLaMA2-7B pre-training model, we have developed MatChat, a groundbreaking large language model explicitly designed for materials science. This model primarily focuses on synthesizing knowledge related to the inorganic materials synthesis process. It can engage in logical reasoning based on the queried materials formula and provides answers in the format of the training set, including formulas, temperature, time, environment conditions, and other relevant information. To facilitate the usage of MatChat, we have further developed a dialogue platform for users based on this model. This platform is currently accessible online at http://chat.aicnic.cn/onchat and is open to researchers in the materials field. This work is poised to inspire and bring new innovative ideas in the field of materials science.

MatChat represents a pioneering effort in the applications of large models within the field of materials science. It currently only supports English languages due to the lack of text data in other languages within the training set. The accuracy of its responses is an area we aim to further refine. The material large language model presented in this study focuses on inorganic chemical synthesis. We aspire for this work to be the 'The Wright brothers' one-minute flight' in the field of inorganic material synthesis pathway prediction. In the future, the research team intends to enhance the model's usability and accuracy by incorporating literature data and information from existing material databases such as Atomly.net, OQMD, etc. Additionally, we plan to optimize the training methodology to enable the large aircraft of inorganic materials synthesis pathway prediction to fly higher and farther.

**Code availability**

The relevant code of this article has been published on the GitHub website, and the link is https://github.com/materialsCnicCas/CASMatChat.

**Acknowledgements**

This work was supported by the Informatization Plan of the Chinese Academy of Sciences, Grant No.CAS-WX2023SF-0101 and XDB33020000, the Key Research Program of Frontier Sciences, CAS, Grant No. ZDBS-LY-7025, and the Youth Innovation Promotion Association CAS (2021167).